\documentclass[prb,eqsecnum,aps,twocolumn,floats,superscriptaddress]{revtex4}
\pdfoutput=1
\usepackage{graphics,dcolumn,float}
\usepackage{graphicx}
\usepackage{dcolumn}
\usepackage{bm}
\usepackage{amsmath}
\usepackage{amssymb}
\usepackage{color}
\begin{document}
\preprint{APS/123-QED}
\title{First Principles Study of Photocatalytic Water Splitting by M$ _{1} $M$ _{2} $CO$ _{2} $ (M$ _{1} $=Zr,Hf;M$ _{2} $=Hf, Ti, Sc)MXenes}

\author{Sima Rastegar}
\affiliation{Department of Physics,
Azarbaijan Shahid Madani University, Tabriz 53714-161, Iran}
\affiliation{Molecular Simulation Laboratory (MSL), Azarbaijan Shahid Madani University, Tabriz, Iran}
\author{Alireza Rastkar Ebrahimzadeh}
\affiliation{Department of Physics,
Azarbaijan Shahid Madani University, Tabriz 53714-161, Iran}
\affiliation{Molecular Simulation Laboratory (MSL), Azarbaijan Shahid Madani University, Tabriz, Iran}
\author{Jaber Jahanbin Sardroodi}
\email[Corresponding author's Email: ]{jsardroodi@azaruniv.ac.ir}
\affiliation{Molecular Simulation Laboratory (MSL), Azarbaijan Shahid Madani University, Tabriz, Iran}
\affiliation{Department of Chemistry,
Azarbaijan Shahid Madani University, Tabriz 53714-161, Iran}

\date{\today}
\begin{abstract}
Using density functional theory (DFT), we investigated the structural, electronic and optical properties of functionalized and doped MXenes such as  M$ _{1} $M$ _{2} $CO$ _{2} $(M$ _{1} $=Zr,Hf;M$ _{2} $=Hf, Ti, Sc). This study aimed to find a suitable photocatalyst that would work well in the water splitting process. Among the calculated nanostructures, MXenes ZrHfCO$ _{2} $ and ZrTiCO$ _{2} $ were chosen as the suitable photocatalysts for the water splitting process. The calculated value of the band gaps with the GGA-PBE functional was 1.08(0.79) eV for the ZrHfCO$ _{2} $ (ZrTiCO$ _{2} $) monolayer. Also, the band gaps for these monolayers with the HSE06 hybrid functional were 1.86 and 1.57 eV, respectively. These MXenes' optical properties, such as complex dielectric function, refractive index, extinction coefficient and reflectivity, were also investigated. The results showed that these monolayers had good absorption in the visible and ultraviolet regions. Additionally, we discovered that ZrHfCO$ _{2} $ and ZrTiCO$ _{2} $ MXenes could be used for the water splitting process by calculating the photocatalytic properties. Meanwhile, the results showed that the monolayers of M$ _{1} $M$ _{2} $CO$ _{2} $ could be promising candidates for photocatalytic, solar energy and optoelectronic applications.
\end{abstract}

%\pacs{
%   77.22.-d 	%Dielectric properties of solids and liquids
%   71.45.Gm 	%Exchange, correlation, dielectric and magnetic response functions, plasmons
%   73.22.-f %	Electronic structure of nanoscale materials and related systems
%   } 

\maketitle

\section{INTRODUCTION}
One of the most essential scientific and technological issues facing humanity in the twenty-first century is providing safe, clean and long-term energy. Currently, fossil fuels make up around 85\% of all energy used worldwide. On the other hand, using fossil fuels would result in enormous greenhouse gas emissions, mostly carbon dioxide, and the global mean temperature rise, which would have permanent consequences on the climate \cite{fujishima1972electrochemical}. Therefore, hydrogen has been proposed as the primary renewable energy carrier; this approach is called the hydrogen economy \cite{borgarello1981photochemical,wang2009metal}.

Since Fujishima and Honda \cite{fujishima1972electrochemical} first published their research on the photoelectrochemical generation of H$ _{2} $ by water splitting, photocatalytic processes have received much attention. Nowadays, photocatalytic processes are regarded as a dramatic and exceptional technology for the direct, low-cost harvesting of clean energy. Many materials are effective in photocatalytic activities under diverse conditions \cite{borgarello1981photochemical,wang2009metal,xiang2012synergetic,chang2014mos2,frame2010cdse,zuo2010self,kudo1989nickel,kato2003highly,maeda2005gan}.

Hydrogen produced by water splitting, using a semiconductor photocatalyst under sunlight, is another energy resource that can replace fossil fuels and serve as a promising solution to important environmental issues \cite{simon2014redox,chen2011increasing,asahi2014nitrogen,ran2014earth}. The photocatalyst must produce electrons and holes to perform the photocatalytic water splitting process. This process can be done via the absorption of solar energy, migration of the generated electrons and holes to the semiconductor surface and the redox reaction of water to form H$ _{2} $ and O$ _{2} $ on the surface.

Meanwhile, to have a suitable and efficient photocatalyst, the following two conditions are necessary:
\\

a)	One of these requirements is that the valence band maximum (VBM) be lower (more positive) than the water oxidation potential (H$ _{2} $O/O$ _{2} $) and the conduction band minimum (CBM) be higher (more negative) than the hydrogen reduction potential (H$ ^{+} $/H$ _{2} $) (Fig. 1)\cite{singh2015computational}.
\\

b)	The second requirement is that the smallest band gap energy required for a semiconductor used as a photocatalyst be 1.23 eV (Fig. 1). 
\\

\begin{figure}[h]%
\centering
\includegraphics[width=0.5\textwidth]{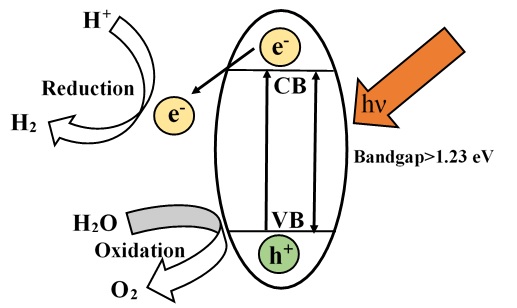}
\caption{Schematic illustration of the main process steps in water splitting.}\label{fig1}
\end{figure}

A novel family of layered materials, namely, 2D transition metal carbides/ nitrides, was effectively separated into primary MX layers from bulk MAX phases using the micromechanical exfoliation process to create graphene \cite{lei2015recent,zhang2016mnpse3}. Because of their chemical and physical features, two-dimensional materials are frequently used in energy storage and material science. Among the unique two-dimensional materials successfully synthesized from the MAX phase, monolayer transition metal carbides such as Ti$ _{2} $C, Ti$ _{3} $C$ _{2} $, V$ _{2} $C, Nb$ _{2} $C and Ta$ _{4} $C$ _{3} $ can be noted \cite{naguib2011two,naguib2012two,naguib2013new,mashtalir2013kinetics}. MXenes are a new monolayer material created by the selective etching of the A layers in the MAX phases \cite{ivanovskii2013graphene,naguib2014two}. MXenes are transition metal carbides and nitrides that have a two-dimensional structure. MXenes \cite{frame2010cdse}, a large class of two-dimensional materials discovered in 2011 that are not graphene. Because of their outstanding stability, electrical conductivity, capacitive nature and redox-based energy storage capacities, these materials are good candidates for energy storage, energy conversion, catalysis and device applications \cite{naguib2012two,xie2014prediction,lukatskaya2016multidimensional,anasori20172d,zhang2018mxene}. 
\\

More than 20 MXenes have been synthesized using transition metals and atomic layer counts. Fig. 2 shows the 2D MXene monolayers of the sort M$ _{2} $C. In these monolayers, functional groups like F, OH and O are spontaneously bound to both sides of the bare MXene from the etchant during the entire etching process, leading to the emergence of the terminated MXene \cite{naguib2014two}. A terminated MXene possesses metallic properties in general. Therefore, band gap engineering is an important technology for using different kinds of MXenes to design new materials and devices for semiconducting, optoelectronic and optical applications \cite{ortega2014band,lin2013indirect}.

Monolayer MXene has been studied in various fields, such as electrochemical energy, storage materials, and hydrogen storage mediums \cite{tang2012mxenes,gan2013first,hu2013mxene,halim2014transparent,ma2014tunable,hu2014two,er2014ti3c2}. With such applications, using monolayer MXene will lead to varied choices for the design of promising nanodevices in optoelectronics and optics out of the pre-existing materials.

\begin{figure}[h]%
\centering
\includegraphics[width=0.5\textwidth]{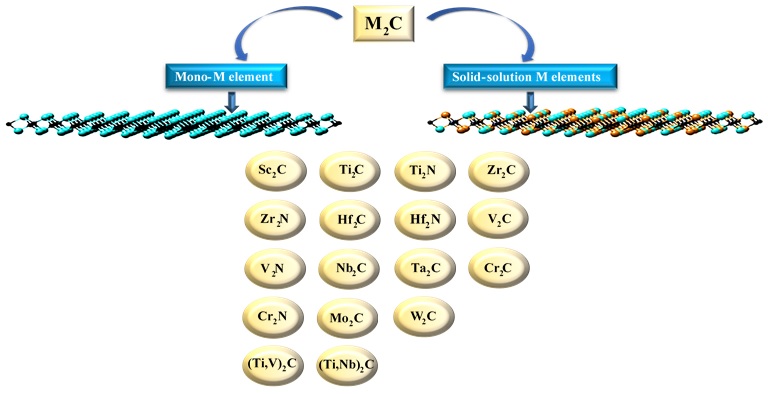}
\caption{A schematic representation of 2D MXene monolayers of sort M$ _{2} $C.}\label{fig2}
\end{figure}

Meanwhile, we have used first-principles density-functional theory (DFT) calculations in this study. More specifically, this paper looks for suitable photocatalysts for the water splitting process using efficient and computational methods. Our findings show that 2D ZrHfCO$ _{2} $ and ZrTiCO$ _{2} $ MXenes are the ideal photocatalysts for water splitting.

\section{Computational methods}
We used the projector augmented plane-wave (PAW) approach, as implemented in the Vienna Ab-initio Simulation Package (VASP) \cite{Kresse1996,Kresse1999,hafner2008ab}, to execute all Density Functional Theory computations. The exchange-correlation functional for structural relaxation was characterized using the generalized gradient approximation (GGA) \cite{Perdew1992} and the Perdew-Burke-Ernzerhof (PBE) \cite{Perdew1996B}. The plane wave cut-off was set to 520 eV. The convergence limit was set as 10$ ^{-6} $ eV and 0.005 eV/$ \mathring{A}$ for energy and force, respectively. The k-points of $ 11\times11\times1 $ and $ 13\times13\times1 $ were employed for geometry optimization, and static self-consistent computations were provided by the Monkhorst-Pack scheme \cite{Monkhorst1976}. The convergence tests were then used to determine the optimal selection of the parameters for the executed calculations based on the system's total energy. In the z-direction, a vacuum layer of $ 20 \AA$ was added to eliminate the interaction between the adjacent images. 
\\

We applied a Heyd-Scuseria-Ernzerhof (HSE06) \cite{heyd2003hybrid} hybrid density functional to compute the electronic structures because the band gaps in the computations with the PBE functional were underestimated. It has been demonstrated that this hybrid functional gives precise values that are in good agreement with the experiments in a wide range of systems, like monolayer MoS$ _{2} $, phosphorene, SrZrO$ _{3} $, TiS$ _{3} $, ZrS$ _{3} $ and HfS$ _{3} $ \cite{liao2014design,sa2014strain,guo2014band,li2015tuning}. Hence, we applied HSE06 to determine the electronic structures and band gap ($ E_{g} $) of the GGA-PBE optimized stages. Also, we employed the HSE06 hybrid functional with an $ 11\times11\times1$ k-point mesh to calculate dielectric constants ($ \epsilon $) at a specific frequency. Finally, the VESTA code was used to analyze the atomic structures.

\section{Results and Discussion}\label{sec2}

\subsection{Structural and electronic properties of 2D M$ _{1} $M$ _{2} $CO$ _{2} $ (M1=Zr,Hf;M2=Hf,Ti,Sc)}
The total energies of the optimized structures were computed at various unit cell volumes to reach the equilibrium lattice constants of the M$ _{1} $M$ _{2} $CO$ _{2} $ monolayer. As shown in Fig. 3, all the suggested configurations shared similar hexagonal structures; it could be seen that 2D ZrHfCO$ _{2} $ and ZrTiCO$ _{2} $, as well as ZrScCO$ _{2} $ and HfScCO$ _{2} $, had the same configuration. Also,  both top and side views for these monolayers are demonstrated in this Figure. The optimized lattice parameters computed by GGA-PBE and the bond length for these monolayers are summarized in Tables I and II, respectively. Besides this, the bond angles were $ \alpha=\beta=90^{\circ} $ and $ \gamma=120^{\circ} $. The obtained results were in good agreement with the other available data\cite{sima2021mxene}.

\begin{figure}[h]%
\centering
\includegraphics[width=0.45\textwidth]{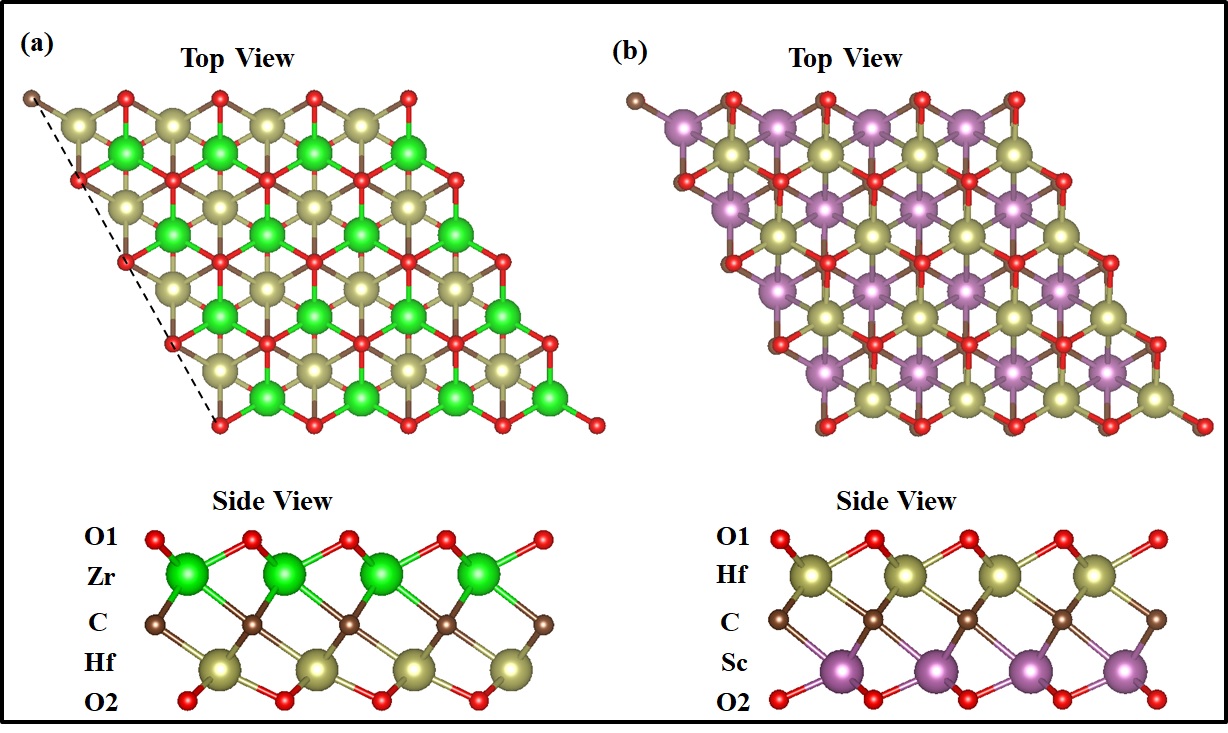}
\caption{Top and side views of the most stable configurations for 2D (a) ZrHfCO$ _{2} $ and (b) HfScCO$ _{2} $.}\label{fig3}
\end{figure}

A useful tool for investigating materials and their applications in the semiconductor industry is the electronic band structure. The band gap in semiconductor materials causes them to have significant optical properties. In the first Brillouin zone (1$ ^{st} $ BZ), the electronic band structures of four 2D MXenes are estimated along the high symmetry directions ($ \Gamma $ M K $ \Gamma $), as illustrated in Fig. 4 and Fig. 5 for PBE and HSE06 functionals. As clearly shown in this Figures, when we perform calculations with PBE and HSE06 functionals, 2D ZrHfCO$ _{2} $ and ZrTiCO$ _{2} $ monolayers are indirect and direct bandgap semiconductors, respectively, with the bandgap from $ \Gamma $ to M. As noted before, the bandgap was estimated using both PBE and HSE06 methods for the computational techniques. Notably, the exchange functionals' self-interaction error may be a reason for underestimating the band gap in PBE. Compared to the PBE functional, the HSE06 hybrid functional is anticipated to produce better bandgap results. The results of calculations illustrated that ZrHfCO$ _{2} $ and ZrTiCO$ _{2} $ monolayers were semiconductors having band gaps of 1.08(0.79) eV and 1.86(1.57) eV with PBE and HSE06, respectively. On the other hand, 2D ZrScCO$ _{2} $ and HfScCO$ _{2} $ monolayers with both PBE and HSE06 functionals were metallic.
\\
\begin{figure*}[h]%
\centering
\includegraphics[width=0.7\textwidth]{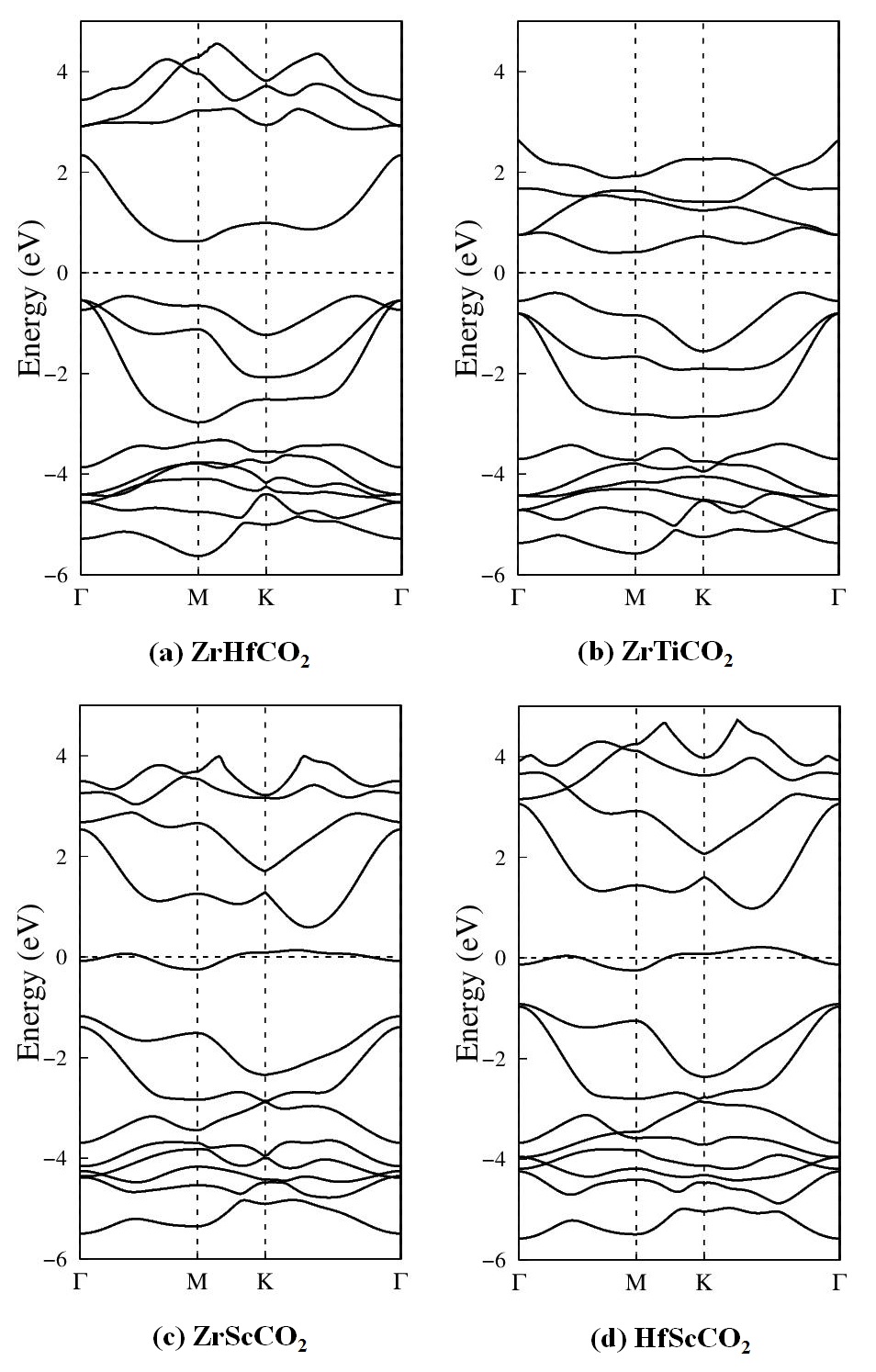}
\caption{Band structures of the functionalized MXenes with geometry M$ _{1} $M$ _{2} $CO$ _{2} $ (M$ _{1} $=Zr, Hf; M$ _{2} $=Hf, Sc, Ti) with GGA-PBE functional. Black dashed lines represent the Fermi level at 0 eV.}\label{fig4}
\end{figure*}

\begin{figure*}[h]%
\centering
\includegraphics[width=0.7\textwidth]{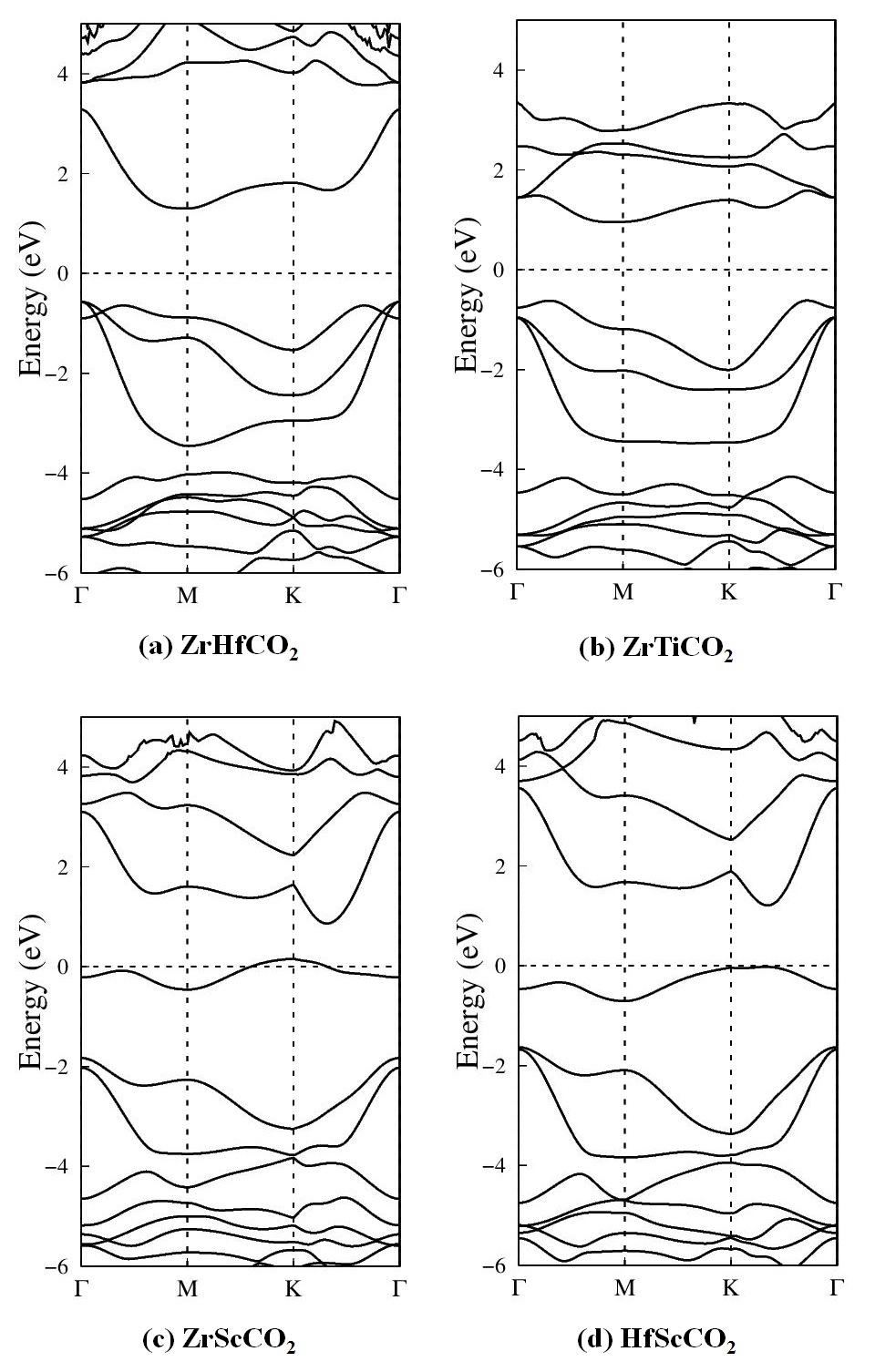}
\caption{Band structures of the functionalized MXenes with geometry M$ _{1} $M$ _{2} $CO$ _{2} $ (M$ _{1} $=Zr, Hf; M$ _{2} $=Hf, Sc, Ti) with HSE06 hybrid functional. Black dashed lines represent the Fermi level at 0 eV.}\label{fig5}
\end{figure*}

Fig. 6 and Fig. 7 display the computed partial electronic density of states for all materials mentioned above for PBE and HSE06 functionals. In ZrHfCO$ _{2} $ and ZrTiCO$ _{2} $ monolayers, the analysis of the partial electronic density of states revealed that the valance band close to the Fermi level was mostly made of the C-p states with just a little overlap with Zr-d, Hf-d and Ti-d orbitals, thus demonstrating the covalent nature of the bonds. Meanwhile, the p or d states were outstanding in the conduction band above the Fermi level.

\begin{figure*}[h]%
\centering
\includegraphics[width=0.5\textwidth]{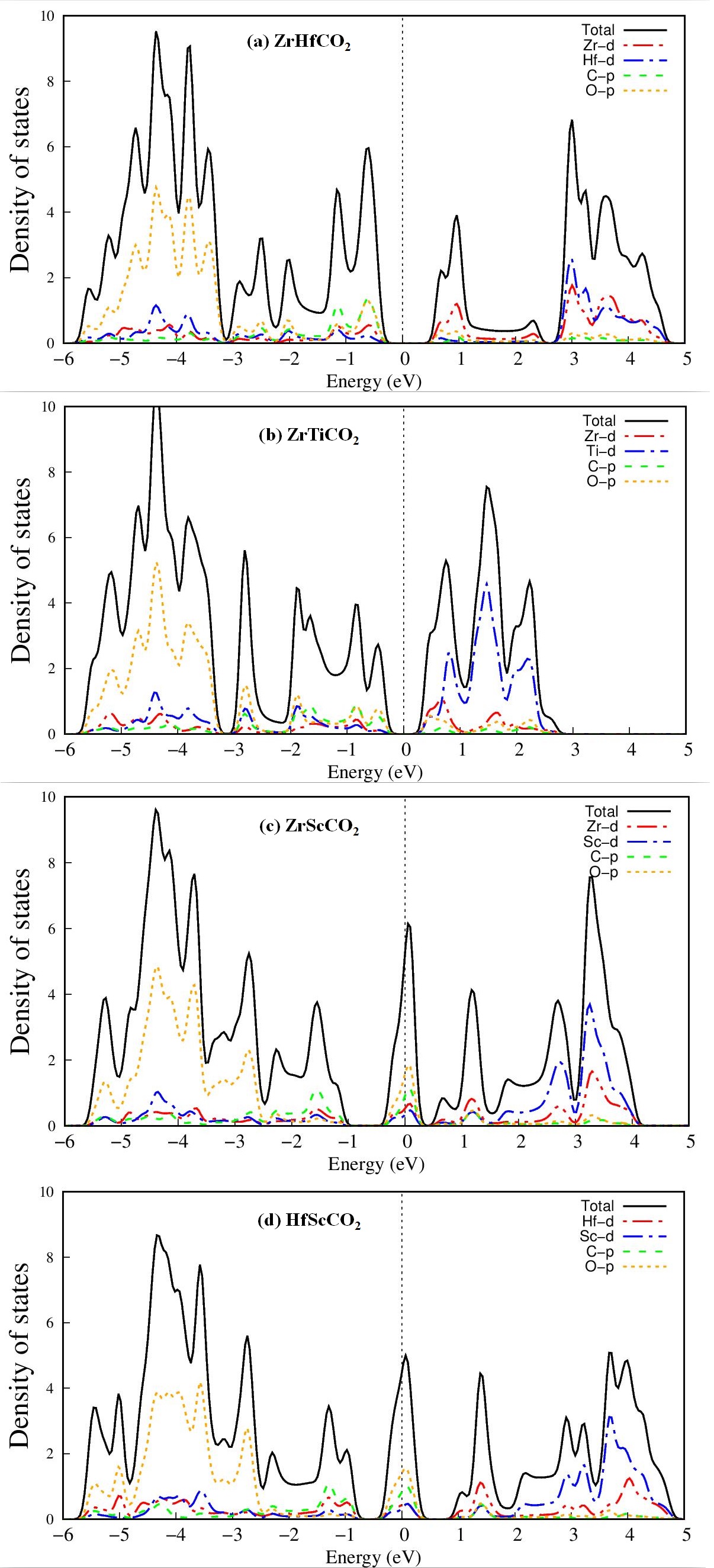}
\caption{Calculated (using GGA-PBE functional) DOS PDOS of 2D (a) ZrHfCO$ _{2} $, (b) ZrTiCO$ _{2} $, (c) ZrScCO$ _{2} $ and (d) HfScCO$ _{2} $. The Fermi energy is set to 0 eV and indicated by the vertical dashed line.}\label{fig6}
\end{figure*}

\begin{figure*}[h]%
\centering
\includegraphics[width=0.5\textwidth]{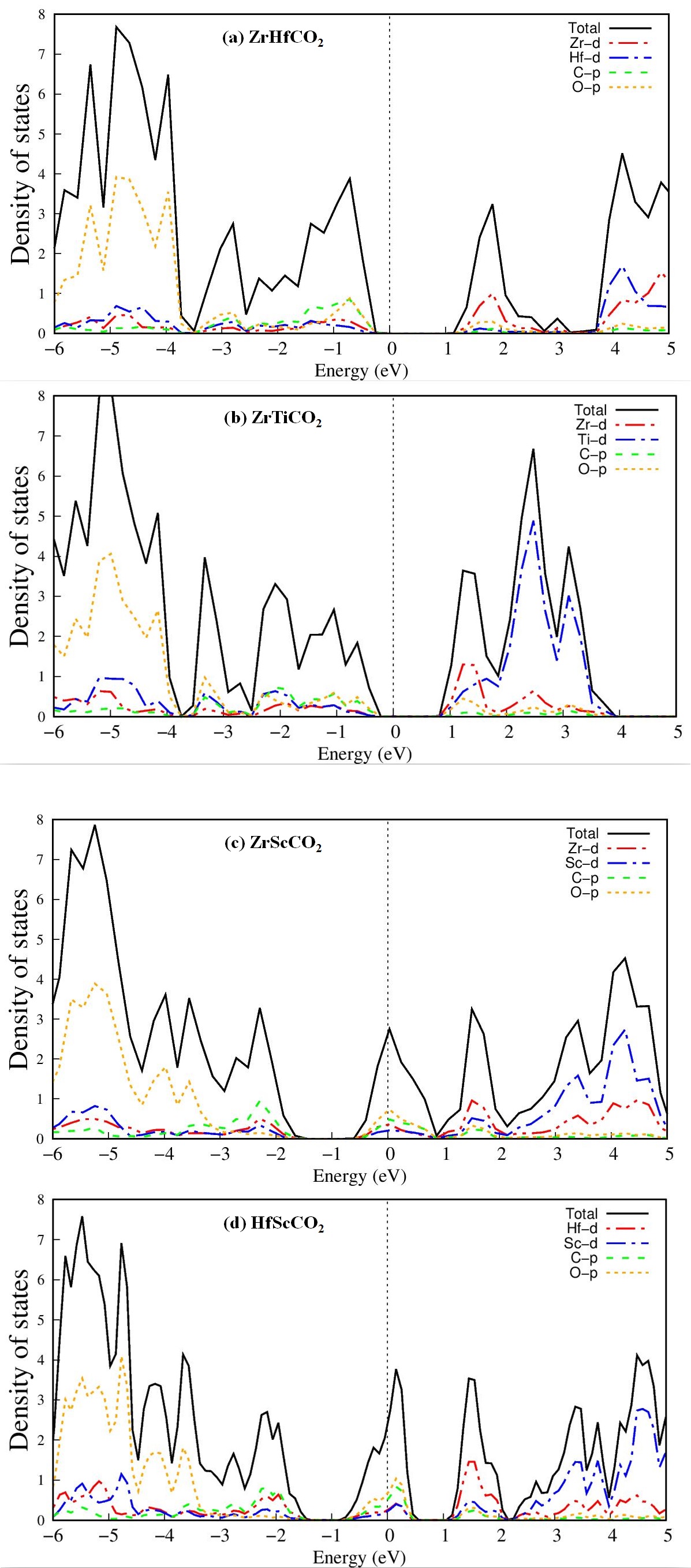}
\caption{Calculated (using HSE06 hybrid functional) DOS PDOS of 2D (a) ZrHfCO$ _{2} $, (b) ZrTiCO$ _{2} $, (c) ZrScCO$ _{2} $ and (d) HfScCO$ _{2} $. The Fermi energy is set to 0 eV and indicated by the vertical dashed line.}\label{fig7}
\end{figure*}

\subsection{Optical properties}
Optical analysis of nanostructures is crucial in nanotechnology, optoelectronics and device management. The optical spectrum provides a huge source of information for studying the band structure, electronic properties and excitations. One of the most important optical features is the complex dielectric function, expressed by the following equation.

\begin{equation}
\varepsilon (\omega) = \varepsilon_{1}(\omega) + i\varepsilon_{2}(\omega),\label{eq1}
\end{equation}

, where $ \epsilon_1(\omega) $ and $ \epsilon_2(\omega) $ are the real and imaginary portions of the dielectric constant, respectively. 
\\

The Kramers Kronig relations can be used to compute the real portion, and the momentum matrix element between the occupied and unoccupied states can be applied to get the imaginary part of the dielectric function. Therefore, the real and imaginary parts of the dielectric function are as follows \cite{abt1994optical}:

\begin{equation}
\varepsilon_{1}^{\alpha \beta} (\omega)= \delta_{\alpha \beta} + \frac{2}{\pi} Pr. \int\limits_{0}^{\infty} \frac{\omega^{\prime}\varepsilon_{2}^{\alpha \beta}(\omega^{\prime})}{(\omega^{\prime})^{2} - (\omega)^{2}}d\omega^{\prime},\label{eq2}
\end{equation}

, where $ Pr $ stands for the principal value of the integral over $ \omega^{\prime} $.

\begin{align}
\nonumber
\varepsilon_{2}^{\alpha \alpha}(\omega) & = \frac{4\pi e^{2}}{m^{2} \omega^{2}}\sum\limits_{i,f}\int \frac{2d^{3}k}{(2\pi)^{3}}\{|\langle ik|P_{\alpha}|fk\rangle|^{2}
\\
& f_{i}^{k}(1-f_{f}^{k})\delta(E_{f}^{k} - E_{i}^{k} - \hbar \omega)\},\label{eq3}
\end{align}

, where $ i $ and $ f $ demonstrate the initial and final state and $ E_{i}^{k} $ is corresponding eigenvalue, $ f_{i}^{k} $ denotes the Fermi distribution, and $ P_{\alpha} $ is a component of the momentum operator. The dielectric function has two intra-band and inter-band portions. We did not consider indirect interband transitions that have little contribution to the optical properties of the material.

We calculated the dielectric constants ($ \epsilon $) at a specific frequency to investigate the optical absorption properties of the 2D semiconductor photocatalyst. The results of the real and imaginary parts of the dielectric function for ZrHfCO$ _{2} $ and ZrTiCO$ _{2} $ monolayers, for both parallel and perpendicular polarization of light, are shown in Fig. 8 and Fig. 9, respectively. In calculating the optical absorption properties of these two-dimensional materials, we used the HSE06 functional. From the real contribution of the dielectric function, the static dielectric value, which is the value of the dielectric function at zero energy, could be obtained. The optical absorption properties of 2D ZrHfCO$ _{2} $ and ZrTiCO$ _{2} $ were analyzed according to the computed imaginary portion of the dielectric function.
\\
 
Considering that the visible region was in the energy range of 1.7 to 3.2 eV, there was one peak of the real part of the dielectric function for 2D ZrHfCO$ _{2} $ and ZrTiCO$ _{2} $ monolayers in the x direction of light, as shown in Fig. 8(b). The peak of the mentioned structures occurred at E=0.69, and 1.42 eV, respectively. ZrTiCO$ _{2} $ had more peak in the visible region than ZrHfCO$ _{2} $ did; Furthermore, the peaks of the real part of the dielectric function in the z direction for ZrTiCO$ _{2} $ were in the visible energy range. The energy values of the peaks in the low energy region up to 3.2 eV (infrared and visible regions of the solar spectrum) for both the real and imaginary parts of the dielectric function are summarized in Table III.
\\

\begin{figure*}[h]%
\centering
\includegraphics[width=0.9\textwidth]{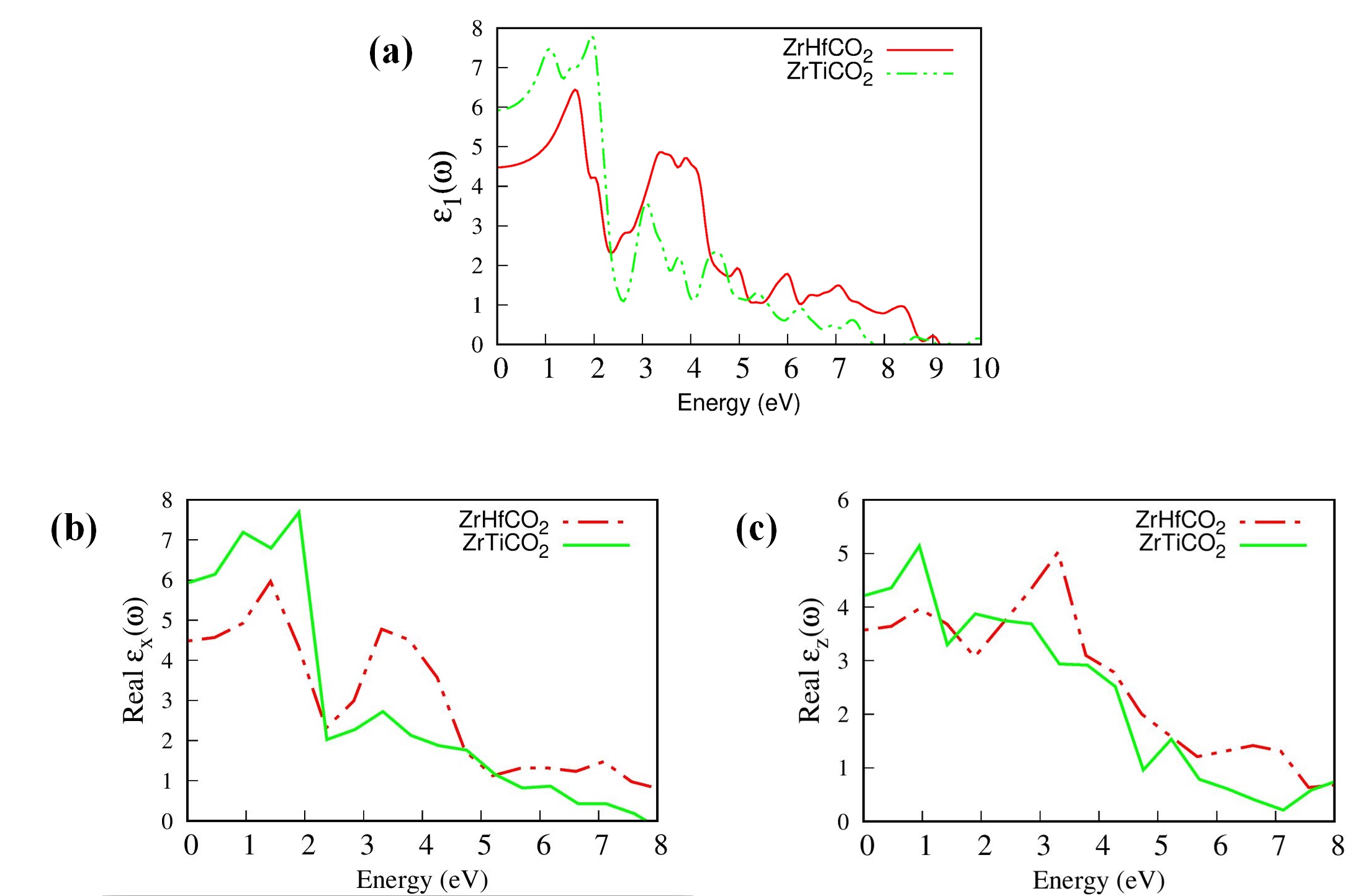}
\caption{The real part of the dielectric function for ZrHfCO$ _{2} $ and ZrTiCO$ _{2} $ monolayers in (a) total state and both light polarization directions, (b) $ E\|x $, and (c) $ E\|z $.}\label{fig8}
\end{figure*}

\begin{figure*}[h]%
\centering
\includegraphics[width=0.9\textwidth]{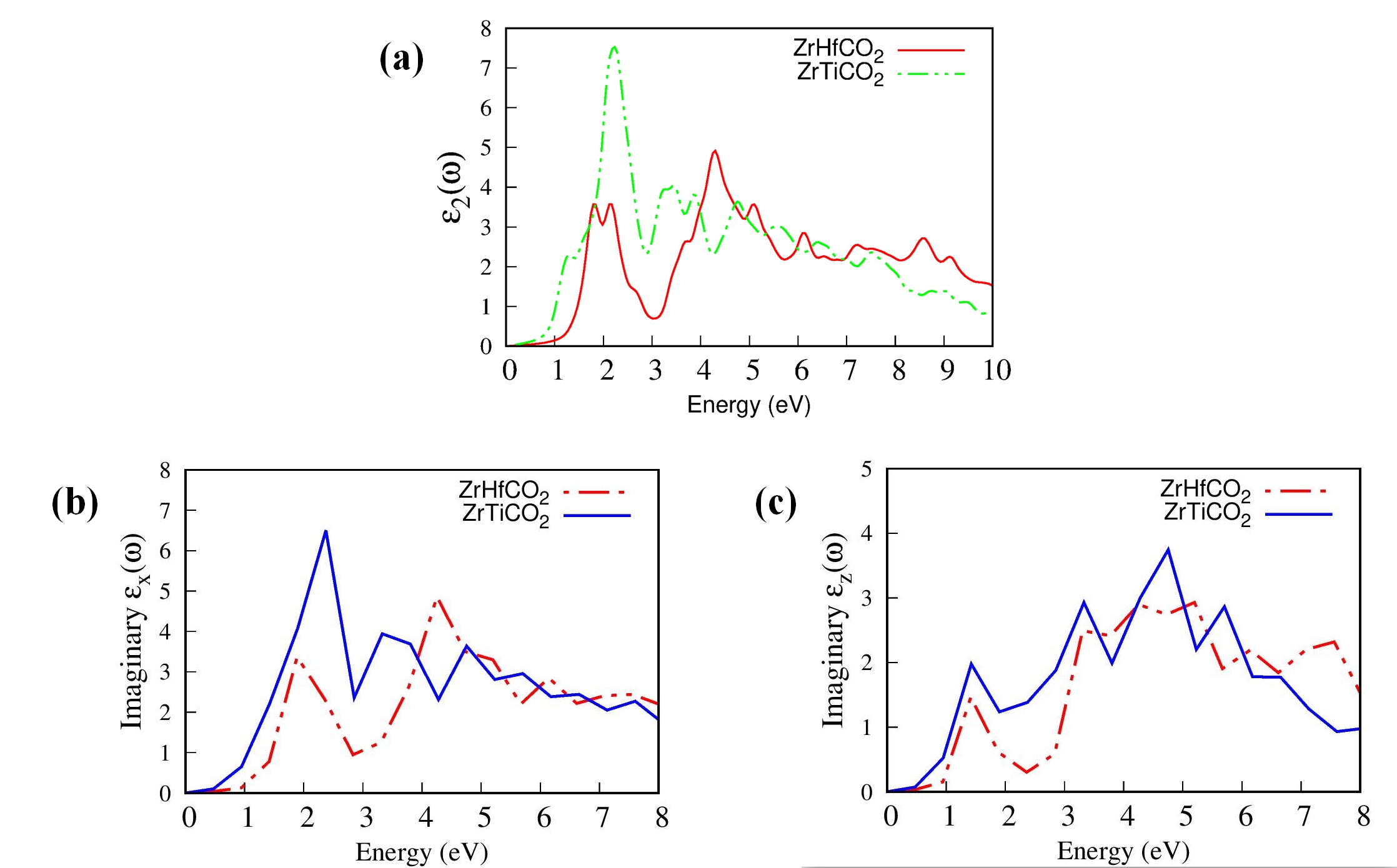}
\caption{The imaginary part of the dielectric function for ZrHfCO$ _{2} $ and ZrTiCO$ _{2} $ monolayers in (a) total state and both light polarization directions, (b) $ E\|x $, and (c) $ E\|z $.}\label{fig9}
\end{figure*}

The transition between the occupied and unoccupied states can be attributed to the imaginary portion of the dielectric function. The onset of absorption is directly related to the band gap of the materials, as seen in Fig. 9. The calculated imaginary part of the dielectric function has major peaks, as summarized in Table III. 

The refractive index, extinction coefficient and reflectivity percentages were investigated. They could be calculated using the equations 4, 5 and 6, respectively.

\begin{equation}
n(\omega) = \bigg(\frac{\sqrt{\varepsilon_{1}^{2}(\omega) + \varepsilon_{2}^{2}(\omega)} + \varepsilon_{1}(\omega)}{2}\bigg)^{\frac{1}{2}}, \label{eq4}
\end{equation}

\begin{equation}
k(\omega) = \bigg(\frac{\sqrt{\varepsilon_{1}^{2}(\omega) + \varepsilon_{2}^{2}(\omega)} - \varepsilon_{1}(\omega)}{2}\bigg)^{\frac{1}{2}}, \label{eq5}
\end{equation}

\begin{equation}
R(\omega) = \bigg|\frac{\sqrt{\varepsilon(\omega)} - 1}{\sqrt{\varepsilon(\omega)} + 1}\bigg|^{2}. \label{eq6}
\end{equation}

These quantities can be calculated by finding the dielectric function's real and imaginary parts. The refractive index, extinction coefficient and reflectivity are shown in Figs. 10(a), 10(b) and 11, respectively. As depicted in Fig. 11, the reflectivity percentage peaks agreed with the real part of the dielectric function (see Fig. 8). The value of reflectivity percentage for 2D ZrTiCO$ _{2} $ was more than that of the ZrHfCO$ _{2} $ monolayer up to E =1.1 eV. Increasing the reflection could reduce the absorption and transmission of electromagnetic waves. ZrHfCO$ _{2} $ and ZrTiCO$ _{2} $ monolayers could not absorb light below 0.04 eV. The M elements and surface termination during the experimental preparation of MXenes determined the optical properties of 2D MXenes. Therefore, by controlling these factors, it is possible to change the optical properties of MXenes.

\begin{figure*}[h]%
\centering
\includegraphics[width=0.8\textwidth]{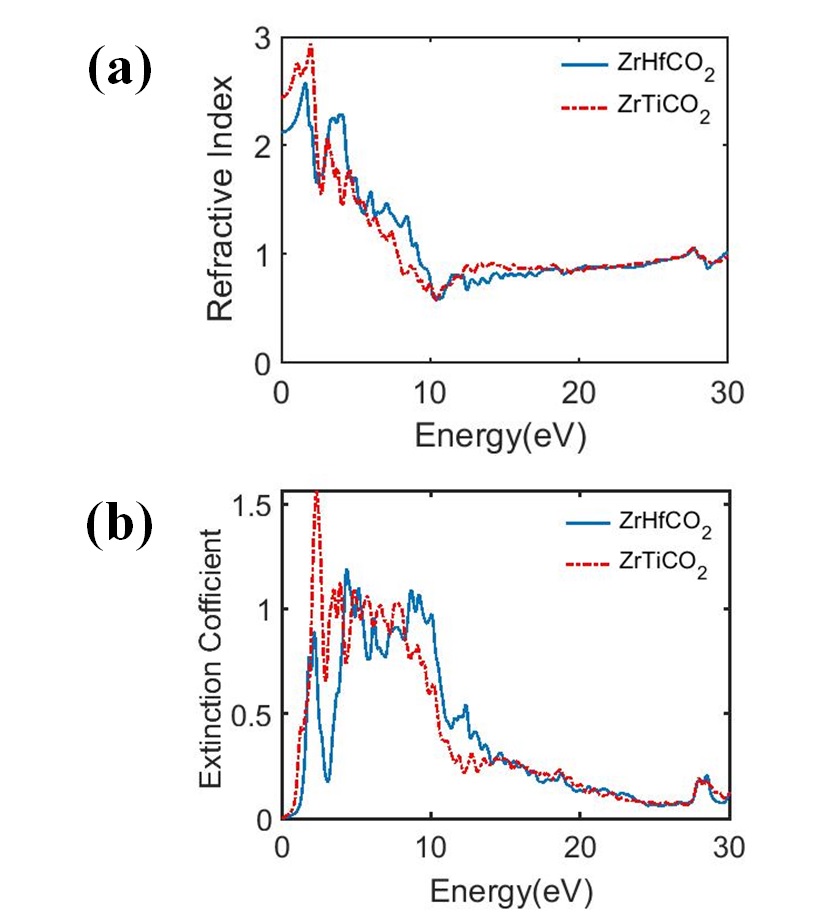}
\caption{The (a) refractive index and (b) extinction coefficient of ZrHfCO$ _{2} $ and ZrTiCO$ _{2} $ from the HSE06 method.}\label{fig10}
\end{figure*}

\begin{figure*}[h]%
\centering
\includegraphics[width=0.7\textwidth]{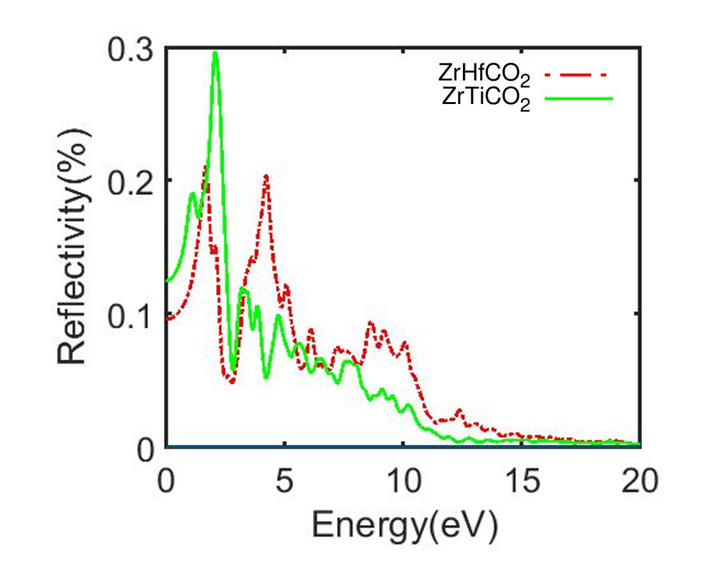}
\caption{The reflectivity percentages of ZrHfCO$ _{2} $ and ZrTiCO$ _{2} $ from the HSE06 method.}\label{fig11}
\end{figure*}

\subsection{Photocatalyst properties}
To obtain and discover a suitable and efficient photocatalyst from MXenes for use in the water splitting process, our simulation calculations consisted of four nanostructures including ZrHfCO$ _{2} $, ZrTiCO$ _{2} $, ZrScCO$ _{2} $ and HfScCO$ _{2} $. The results of calculations illustrated that the two pristine MXenes ZrScCO$ _{2} $ and HfScCO$ _{2} $ were metallic. In contrast, two functionalized M$ _{1} $M$ _{2} $C-type MXenes, ZrHfCO$ _{2} $ and ZrTiCO$ _{2} $, were semiconductors with the band gaps of 1.08(1.86) and 0.79(1.57) eV and GGA-PBE (HSE06) functionals, respectively. Hence, we focus on ZrHfCO$ _{2} $ and ZrTiCO$ _{2} $ to use them as photocatalysts in the water splitting process.
\\

An incident photon on a semiconductor with energy greater than $ E_{g} $ can produce electron-hole pairs. The excited electrons (holes) can participate in redox processes, as well as the possibility of recombination of photogenerated electron-hole pairs. One of the requirements for a suitable water splitting photocatalyst is that the valence band maximum (VBM) be lower (more positive) than the water oxidation potential (H$ _{2} $O/O$ _{2} $) and the conduction band minimum (CBM) be higher (more negative) than the hydrogen reduction potential (H$ ^{+} $/H$ _{2} $). The level of CBM ($ E_{CBM} $) and VBM ($ E_{VBM} $) energies is computed by the following formulas to verify these conditions \cite{katsumata2014z}:

\begin{equation}
E_{VBM} = X - E_{e} + \frac{1}{2}E_{g}, \label{eq7}
\end{equation}

\begin{equation}
E_{CBM} = E_{VBM} - E_{g}. \label{eq8}
\end{equation}

Here, X is the semiconductor's electronegativity, and $ E_{e} $ is the energy of free electrons on the hydrogen scale (-4.5 eV). The calculated value of X was 5.16 eV (5.17 eV) for the ZrHfCO$ _{2} $ (ZrTiCO$ _{2} $) monolayer.  Fig. 12 represents the obtained VBM and CBM edge potentials. According to the results, 2D ZrHfCO$ _{2} $ and ZrTiCO$ _{2} $ could be proper photocatalyst candidates for the O$ _{2} $/H$ _{2} $O water splitting half reaction. The calculated $ E_{VBM} $ and $ E_{CBM} $ of the structures within the HSE06 method are presented in Table III.

\begin{figure*}[h]%
\centering
\includegraphics[width=0.7\textwidth]{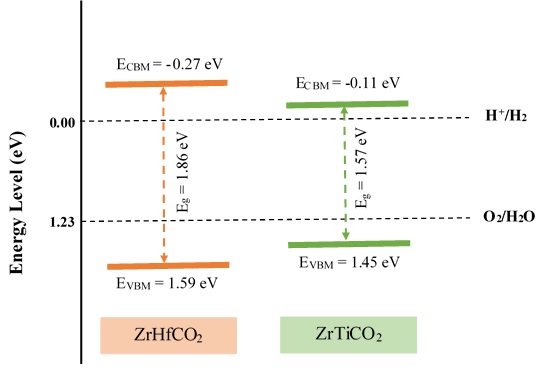}
\caption{Energy diagrams of the VBM and CBM edge potential of ZrHfCO$ _{2} $ and ZrTiCO$ _{2} $ monolayers for the normal hydrogen electrode (NHE).}\label{fig12}
\end{figure*}

\begin{table*}[th]%[!hbtp]
\centering
\caption{The calculated lattice parameter (a) for 2D ZrHfCO$_{2}$, ZrTiCO$_{2}$, ZrScCO$_{2}$ and HfScCO$_{2}$ MXenes}
\begin{tabular}{cccccc}
\hline
Lattice parameter(a) \qquad \qquad &  ZrHfCO$_{2}$ & \qquad ZrTiCO$_{2}$ & \qquad  ZrScCO$_{2}$ & \qquad HfScCO$_{2}$\\\hline
GGA-PBE \qquad \qquad  & 3.3120 & \qquad 3.2863 & \qquad 3.2897 & \qquad 3.2206\\
\end{tabular}
\label{tab:compareFreq}
\end{table*}

\begin{table*}[th]%[!hbtp]
\centering
\caption{Energy gap and bond length for 2D ZrHfCO$_{2}$, ZrTiCO$_{2}$, ZrScCO$_{2}$, and HfScCO$_{2}$ MXenes}
\begin{tabular}{cccccc}
\hline
 \qquad \qquad &  ZrHfCO$_{2}$ & \qquad ZrTiCO$_{2}$ & \qquad   ZrScCO$_{2}$ & \qquad  HfScCO$_{2}$ & \\\hline
E$_{g}$(GGA-PBE) \qquad \qquad  & 1.08 & \qquad 0.79 & \qquad $-$ & \qquad $-$ & \\\hline
E$_{g}$(HSE06) \qquad \qquad  & 1.86 & \qquad 1.57 & \qquad $-$ & \qquad $-$ & \\\hline 
M$ _{1} $-M$ _{2} $ \qquad \qquad  & 3.41 & \qquad $ 3.31 $ & \qquad $ 3.41 $ & \qquad $ 3.39 $ & \\\hline
M$ _{1} $-C \qquad \qquad  & $ 2.43 $ & \qquad 2.40 & \qquad $ 2.33 $ & \qquad $ 2.32 $ & \\\hline
M$ _{1} $-O \qquad \qquad  & $ 2.16 $ & \qquad $ 2.16 $ & \qquad 2.10 & \qquad $ 2.17 $ & \\\hline
M$ _{2} $-C \qquad \qquad  & $ 2.33 $ & \qquad $ 2.27 $ & \qquad $ 2.41 $ & \qquad $ 2.42 $ & \\\hline
M$ _{2} $-O \qquad \qquad  & 2.12 & \qquad 2.06 & \qquad 2.06 & \qquad $ 2.05 $ & \\\hline
 \qquad \qquad  &  & \qquad  & \qquad  & \qquad  & \qquad \\
\end{tabular}
\label{tab:compareFreq}
\end{table*}

\begin{table*}[th]%[!hbtp]
\centering
\caption{The calculated peak energies along x- and z-directions, E$ _{VBM} $ and E$ _{CBM} $ for 2D ZrHfCO$ _{2} $ and ZrTiCO$ _{2} $ monolayers by HSE06 method.}
\begin{tabular}{ccccccc}
\hline
 \qquad \qquad &  $ \Re. E\lVert x(eV) $ &  \qquad  $ \Re. E\lVert z(eV) $ & \qquad  $ \Im. E\lVert x(eV) $ & \qquad $ \Im. E\lVert z(eV) $ & \qquad  E$ _{VBM}(eV) $ & \qquad  E$ _{CBM}(eV) $ \\\hline
ZrHfCO$ _{2} $ \qquad \qquad  &  $1.8$ & \qquad $1.1$ & \qquad $1.9$ & \qquad $1.6$ & \qquad 1.59 & \qquad -0.27\\\hline
ZrTiCO$ _{2} $ \qquad \qquad  & $2.1$ & \qquad $2.2$ & \qquad $2.5$ & \qquad $1.8$ & \qquad 1.45 & \qquad -0.11\\\hline \qquad \\
\end{tabular}
\label{tab:compareFreq}
\end{table*}

\section{Conclusion}\label{sec13}
The structural, electronic, optical and photocatalytic properties of 2D M$ _{1} $M$ _{2} $CO$ _{2} $ (M$ _{1} $= Zr, Hf; M$ _{2} $=Hf, Sc, Ti) MXenes were investigated using density functional theory. Our calculations results demonstrated that ZrHfCO$ _{2} $ and ZrTiCO$ _{2} $ monolayers could have semiconducting properties with an indirect band gap of 1.08(1.86) eV and 0.79(1.57) eV and GGA-PBE (HSE06) functional, respectively. Meanwhile, the other two monolayers ZrScCO$ _{2} $ and HfScCO$ _{2} $ were conductive with both functionals. Various optical properties, including the real and imaginary parts of the dielectric function, refractive index, extinction coefficient and reflectivity of ZrHfCO$ _{2} $ and ZrTiCO$ _{2} $ monolayers, were computed. Also, the valence and conduction band edge potentials were calculated to study the photocatalytic properties of these monolayers. It was revealed that ZrHfCO$ _{2} $ and ZrTiCO$ _{2} $ monolayers could be used as photocatalysts for the O$ _{2} $/H$ _{2} $O water splitting half-reaction based on the photocatalytic properties of them. Consequently, semiconducting ZrHfCO$ _{2} $ and ZrTiCO$ _{2} $ MXenes could be effective candidates for optoelectronic devices, electromagnetic interference shielding, and photocatalysis in the future, thanks to the engineering of their electronic, optical and photocatalytic properties.

\bibliographystyle{apsrev4-1}
\bibliography{refrences.bib}
\end{document}